\documentclass[sigconf, authorversion]{acmart}
\AtBeginDocument{%
  }

\usepackage[most]{tcolorbox}

\setcopyright{acmlicensed}
\copyrightyear{2025}
\acmYear{2025}
\acmConference[CIKM'25]{The 34th ACM International Conference on Information and Knowledge Management}{November~10--14, 2025}{Seoul}

\begin{document}

\title{FinAI Data Assistant: LLM-based Financial Database Query Processing with the OpenAI Function Calling API}

\author{Juhyeong Kim}
\email{juhyeong.kim@miraeasset.com}
\email{nonconvexopt@gmail.com}
\affiliation{
  \institution{Mirae Asset Global Investments}
  \institution{AI Quant Lab, MODULABS}
  \city{Seoul}
  \country{Republic of Korea}
}

\author{Yejin Kim}
\email{yejin.kim.ds@meritz.com}
\affiliation{
  \institution{Meritz Fire \& Marine Insurance}
  \institution{AI Quant Lab, MODULABS}
  \city{Seoul}
  \country{Republic of Korea}
}

\author{Youngbin Lee}
\email{youngandbin@elicer.com}
\affiliation{
  \institution{Elice}
  \institution{AI Quant Lab, MODULABS}
  \city{Seoul}
  \country{Republic of Korea}
}

\author{Hyunwoo Byun}
\email{hyunwoo.byun@miraeasset.com}
\affiliation{
  \institution{Mirae Asset Global Investments}
  \city{Seoul}
  \country{Republic of Korea}
}

\renewcommand{\shortauthors}{Kim et al.}

\begin{abstract}
We present \textbf{FinAI Data Assistant}, a practical approach for natural-language querying over financial databases that combines large language models (LLMs) with the OpenAI Function Calling API. Rather than synthesizing complete SQL via text-to-SQL, our system routes user requests to a small library of vetted, parameterized queries, trading generative flexibility for reliability, low latency, and cost efficiency. We empirically study three questions: (RQ1) whether LLMs alone can reliably recall or extrapolate time-dependent financial data without external retrieval; (RQ2) how well LLMs map company names to stock ticker symbols; and (RQ3) whether function calling outperforms text-to-SQL for end-to-end database query processing. Across controlled experiments on prices and fundamentals, LLM-only predictions exhibit non-negligible error and show look-ahead bias primarily for stock prices relative to model knowledge cutoffs. Ticker-mapping accuracy is near-perfect for NASDAQ-100 constituents and high for S\&P~500 firms. Finally, FinAI Data Assistant achieves lower latency and cost and higher reliability than a text-to-SQL baseline on our task suite. We discuss design trade-offs, limitations, and avenues for deployment.
\end{abstract}




\maketitle

\section{Introduction}
Financial analysts, portfolio managers, and operations teams increasingly expect to access structured market and fundamental data using natural language. However, directly answering such queries with an LLM alone remains challenging for three reasons. First, LLMs have knowledge cutoffs and do not observe live data, making temporal accuracy difficult without external retrieval. Second, free-form generation can hallucinate field names, units, or filters, which is problematic in regulated settings where determinism, auditability, and cost control are paramount. Third, generic text-to-SQL pipelines often emit verbose or invalid queries, incurring latency and operational cost while still requiring extensive guardrails.

To address these challenges, we develop \textbf{FinAI Data Assistant}, a system that integrates an LLM with the OpenAI Function Calling API and a small set of vetted, parameterized SQL templates tailored to common financial data tasks (e.g., market prices and firm fundamentals). The LLM performs high-level intent classification and argument extraction; execution is delegated to trusted linking functions that instantiate the appropriate query with validated parameters. This design preserves the usability of natural-language interaction while ensuring stable latency, predictable cost, and reduced surface area for errors.

We organize our study around three research questions (RQs): (\textbf{RQ1}) Is integration with external databases necessary for accurate handling of temporal financial information before and after model knowledge cutoffs? (\textbf{RQ2}) To what extent can OpenAI LLMs correctly map company names to stock ticker symbols? (\textbf{RQ3}) Does the Function Calling approach deliver higher accuracy and efficiency than text-to-SQL for end-to-end database query processing?

Our contributions are threefold. \textit{Design:} We describe a pragmatic architecture that uses a compact, reusable query library instead of generating full SQL from scratch. \textit{Evaluation:} We present controlled experiments that quantify limitations of LLM-only recall, measure ticker-mapping coverage, and compare Function Calling with a text-to-SQL baseline in terms of latency, cost, and reliability. \textit{Findings:} We observe that (i) LLM-only predictions produce non-trivial errors and exhibit look-ahead bias mainly for prices, (ii) ticker mapping is near-perfect for NASDAQ-100 and strong for S\&P~500 names, and (iii) FinAI Data Assistant is consistently faster and cheaper than text-to-SQL on our task suite while maintaining perfect task completion in our cases.

\section{Related Works}

Natural language interfaces to data (NLIDB) have a long history across databases and NLP. Recent surveys synthesize progress on semantic parsing for data access and, more specifically, LLM-based database interfaces \cite{quamar2022nli2d,hong2024nextgen,zhu2024llmtexttosqlsurvey,shi2025llmtexttosqlsurvey}. Core challenges recur throughout: schema linking, compositional generalization, disambiguation under incomplete context, and execution-time robustness under syntactic and semantic constraints.

\textbf{Text-to-SQL.}
Early neural approaches such as Seq2SQL and SQLNet targeted single-table settings (e.g., WikiSQL) via constrained decoding or sketch-based generation \cite{Zhong2017Seq2SQL,Xu2017SQLNet}. The field subsequently shifted to cross-domain, multi-table parsing with the introduction of Spider and CoSQL, catalyzing research on schema encoding and conversational querying \cite{Yu2018Spider,Yu2019CoSQL}. State-of-the-art parsers model rich schema relations and execution constraints (e.g., RAT-SQL, BRIDGE, SmBoP, PICARD), or unify structured grounding tasks in a text-to-text formulation (UnifiedSKG) \cite{Wang2020RATSQL,Lin2020BRIDGE,Rubin2021SmBoP,Scholak2021PICARD,Xie2022UnifiedSKG}. With the advent of LLMs, in-context learning, decomposition, and self-correction further improve robustness (e.g., DIN-SQL, RESDSQL), while surveys provide comprehensive comparisons of prompting and tool-augmented strategies \cite{Pourreza2023DINSQL,Li2023RESDSQL,zhu2024llmtexttosqlsurvey,shi2025llmtexttosqlsurvey}. Despite progress, pure text-to-SQL pipelines can emit lengthy or invalid queries, require extensive guardrails, and incur variable latency and cost in production.

\textbf{Tool use and function calling.}
A complementary line of work studies LLM agents that reason while invoking tools or APIs. ReAct interleaves reasoning and acting, Toolformer explores self-supervised tool-use training, and Gorilla maps natural language to large API surfaces \cite{Yao2023ReAct,Schick2023Toolformer,Prasad2024Gorilla}. In practical systems, function calling constrains the model to produce structured arguments for pre-registered capabilities, enabling determinism and safer integration \cite{OpenAI2023FunctionCalling}. Our approach adopts this paradigm for financial data access: instead of synthesizing arbitrary SQL, we route requests to a compact library of vetted, parameterized queries, balancing usability with auditability, latency, and predictable cost.

\textbf{LLMs in finance and forecasting.}
Domain-specific LLMs (e.g., BloombergGPT) and open initiatives (e.g., FinGPT) investigate financial knowledge and downstream utility \cite{Wu2023BloombergGPT,Yang2023FinGPT}. In time series, LLMs show promising zero-shot behavior \cite{gruver2023zeroshot}, yet recent studies question whether apparent forecasting skill reflects memorization or ex-post information, proposing ex-ante benchmarks and careful evaluation protocols \cite{lopezlira2025memorization,liu2025exante,crane2025totalrecall,pham2024basechatgpt}. Relatedly, prior work highlights look-ahead bias risks in finance-oriented LLM pipelines (e.g., GPT-based sentiment for return prediction) \cite{glasserman2024lookahead}. These findings motivate our RQ1 analysis and our decision to integrate explicit database retrieval rather than rely on model recall of time-varying quantities. Recent studies have actively investigated the portfolio-optimization capabilities of large language models (LLMs) \cite{kim2025guruagents,lee2025llmenhanced}.

\textbf{Evaluation methodology.}
For statistical comparisons under unequal variances and sample sizes, we follow standard practice using Welch’s $t$-test with degrees-of-freedom approximation per Satterthwaite \cite{welch1947generalization,satterthwaite1946approximate}. Our experimental design, however, emphasizes end-to-end metrics (latency, token cost, task completion) that reflect practical system behavior under realistic financial queries.

\section{Methods}
\subsection{Preliminaries: OpenAI Function Calling API}
OpenAI introduced the Function Calling API to allow developers to define functions and have models produce structured outputs (e.g., JSON arguments) that can be executed by external systems. This mechanism bridges natural-language inputs and programmatic actions.
\\
\textbf{Advantages.}
\begin{itemize}
\item \textbf{Structured Output} — Models emit standardized JSON arguments, reducing ambiguity and simplifying downstream processing.
\item \textbf{System Integration} — Natural-language requests can be mapped to function calls that interact with databases, APIs, or computation engines.
\item \textbf{Error Reduction} — Compared with free-form text, structured outputs reduce parsing errors and improve reliability in multi-step pipelines.
\item \textbf{Workflow Automation} — Natural language serves as a high-level interface for automating complex workflows.
\end{itemize}

\textbf{Limitations.}
\begin{itemize}
\item \textbf{Design Overhead} — Developers must carefully specify schemas; poor schemas yield incorrect or incomplete outputs.
\item \textbf{Latency and Cost} — Function invocation introduces overhead relative to pure text generation.
\item \textbf{Security} — As model outputs can trigger actions, validation and sanitization are essential.
\end{itemize}

\textbf{Basic Usage.}
A typical implementation has three stages: (1) \textbf{Function Definition} specifying names, descriptions, and parameter schemas; (2) \textbf{Model Invocation} with both user input and function specs so the model can choose whether/how to call a function; and (3) \textbf{Function Execution} where returned JSON arguments are validated and dispatched to application code. The flow is thus \textit{natural language} $\to$ \textit{structured JSON} $\to$ \textit{function execution}.

\subsection{FinAI Data Assistant}
We propose \textbf{FinAI Data Assistant}, a database query processing method that integrates OpenAI LLMs and Function Calling to answer natural-language questions about financial data in a fast and economical manner. We assume the service provider maintains existing financial databases (e.g., market and fundamental data). Implementation proceeds by defining a \emph{minimal, common set} of parameterized SQL statements for each data type. This reduces redundancy relative to text-to-SQL systems and lowers both cost and error rate.

Concretely, we build separate base statements for market data and fundamental data. We then implement linking functions that execute these statements with validated parameters. Finally, an LLM determines which linking function to use and what parameter values to supply. This design concentrates complexity in a small, auditable query library while preserving the flexibility of natural-language interaction.

\section{Experiments}

\subsection{Necessity of LLM--Database Integration (RQ1)}
We evaluate whether LLMs alone can accurately recover financial time series given only a company name and short histories. Specifically, we ask models to predict stock prices (daily closes) and fundamental metrics (revenue and net income) at specified periods. For stock prices, the model predicts 10 daily values from the preceding 10 business days; for fundamentals, it predicts 8 quarters from the previous 8 quarters. Dates are explicitly provided.

We use the following prompt template:
\begin{tcolorbox}[title=Prompt,
  colback=green!10,   
  colframe=green!40!black, 
  boxrule=0.6pt,       
  arc=2mm,             
  breakable            
]
\ttfamily
You are given the following \{target\} data for the company \{company\_name\}.\\
Predict the next \{num\_output\} values based on the past \{num\_input\} values.\\[4pt]
Instructions:\\
- Do not add any explanations.\\
- Do not include quotation marks, backticks, or annotations.\\
- Return only the predicted values separated by commas (e.g., 1.0, 2.0, 3.0).\\[4pt]
Data:\\
\{dates\}\\
\{values\}
\end{tcolorbox}

For each ticker $i$ with horizon length $T$, we compute the log mean squared error (MSE):
\begin{equation}
\log(\mathrm{MSE}_i) 
= \log\left(\frac{1}{T} \sum_{t=1}^{T} (\hat{x}_{it} - x_{it})^2\right),
\end{equation}
where $\hat{x}_{it}$ and $x_{it}$ denote predictions and ground truth, respectively.

\begin{table}[h]
  \caption{Average $\log(\mathrm{MSE})$ across tickers. Left values use data before each model's knowledge cutoff; right values use the latest data after the cutoff.}
  \label{tab:log_sSSE}
  \centering
  \begin{tabular}{lccc}
    \toprule
    Model & \textbf{Stock Price} & \textbf{Revenue} & \textbf{Net Income} \\
    \midrule
    gpt-4o & $-2.54\,|\,-1.39$  & $6.94\,|\,6.41$ & $6.21\,|\,6.24$ \\
    gpt-4.1 & $-2.13\,|\,-1.33$ & $6.33\,|\,6.18$ & $5.88\,|\,5.83$ \\
    gpt-5 & $-2.18\,|\,-1.05$ & $7.18\,|\,6.63$ & $6.19\,|\,6.16$ \\
    \bottomrule
  \end{tabular}
\end{table}

Table~\ref{tab:log_sSSE} shows that LLM-only predictions incur non-negligible error, indicating that models cannot reliably retrieve exact financial values from short histories without database access.

As a supplementary analysis, we test for look-ahead bias with respect to each model's knowledge cutoff via a one-sided Welch's t-test. \cite{welch1947generalization} Welch's T-test allow us to compare the means of group with different variances. Let
\begin{equation}
X_i = \log(\mathrm{MSE}_i^{\,b}),\ \ i \in G_1,\ \ n_1 = |G_1|
\end{equation}

\begin{equation}
Y_j = \log(\mathrm{MSE}_j^{\,a}),\ \ j \in G_2,\ \ n_2 = |G_2|
\end{equation}
where superscripts $b$ and $a$ denote before and after the knowledge cutoff. The test statistic is
\begin{equation}
 t = \frac{\bar{X} - \bar{Y}}{\sqrt{\frac{s_X^2}{n_1} + \frac{s_Y^2}{n_2}}},
\end{equation}
with degrees of freedom approximated by the Welch--Satterthwaite formula \cite{satterthwaite1946approximate}:
\begin{equation}
\nu = \frac{\left(\tfrac{s_X^2}{n_1} + \tfrac{s_Y^2}{n_2}\right)^2}{\tfrac{(s_X^2/n_1)^2}{n_1 - 1} + \tfrac{(s_Y^2/n_2)^2}{n_2 - 1}}.
\end{equation}
We obtain p-values using the Student's $t$ distribution with $\nu$ degrees of freedom.

\begin{table}[h]
  \caption{P-values for one-sided Welch's t-test.}
  \label{tab:t_test}
  \centering
  \begin{tabular}{lccc}
    \toprule
    Model & \textbf{Stock Price} & \textbf{Revenue} & \textbf{Net Income} \\
    \midrule
    gpt-4o & 0.0003 & 0.8764 & 0.4681 \\
    gpt-4.1 & 0.0137 & 0.6262 & 0.5381 \\
    gpt-5 & 0.0003 & 0.8568 & 0.5252 \\
    \bottomrule
  \end{tabular}
\end{table}

At the 0.05 significance level, only the stock price comparisons reject the null hypothesis, suggesting look-ahead bias is present primarily for stock prices, not for revenue or net income.

\subsection{Basic Company Information Understanding (RQ2)}
\begin{figure*}[t]
  \includegraphics[width=\textwidth]{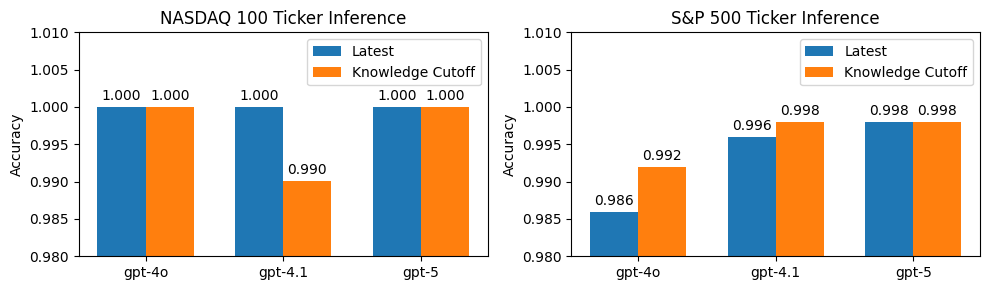}
  \caption{Ticker coverage test. We provide a company name and ask the model for the corresponding ticker. ``Latest'' denotes accuracy for the most recent index constituents. ``Knowledge Cutoff'' denotes accuracy for the same sets restricted to membership before each model's knowledge cutoff.}
  \Description{Ticker coverage is measured by independently asking for each company's ticker given its name, using a consistent prompt.}
  \label{fig:exp2}
\end{figure*}

To address RQ2, we evaluate OpenAI LLMs on a ticker-retrieval task: given a company name, return its stock ticker. Strong performance would indicate that models retain sufficient entity knowledge to support common financial queries. Figure~\ref{fig:exp2} summarizes results.

For NASDAQ-100 constituents, accuracy is effectively perfect for both the latest set and the pre-cutoff set for each model (with a single miss for \texttt{gpt-4.1}). For S\&P~500 companies, accuracy increases with newer models; as expected, performance for the latest constituents is slightly lower than for the pre-cutoff set, but overall accuracy remains high. These results suggest that OpenAI LLMs can reliably map company names to tickers, especially for widely covered firms.

\subsection{Database Query Performance (RQ3)}
\begin{figure*}[t]
  \centering
  \includegraphics[width=\textwidth]{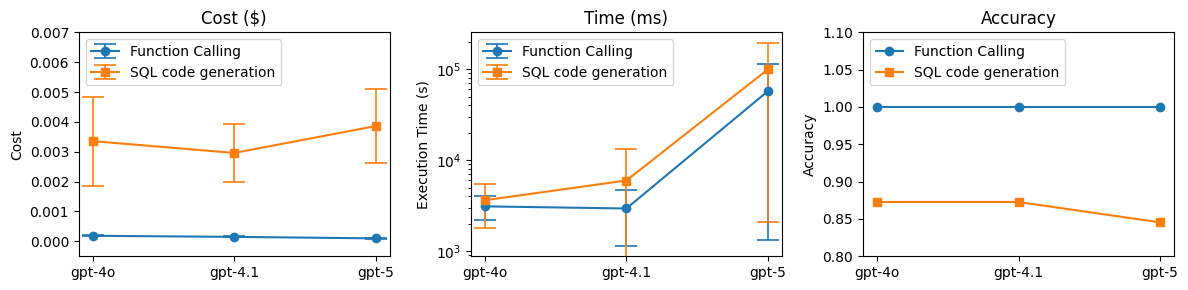}
  \caption{Cost, time, and accuracy for database query processing: Function Calling (FinAI Data Assistant) versus text-to-SQL. All metrics are measured for successful completion of each test case. Error bars in the cost and time plots indicate $\pm 1$ standard deviation.}
  \label{fig:exp3}
  \Description{}
\end{figure*}

Figure~\ref{fig:exp3} compares end-to-end cost, latency, and task accuracy. Blue denotes \textbf{FinAI Data Assistant} (Function Calling); orange denotes the text-to-SQL baseline. A representative task is: ``What are the latest 10 daily closing prices for NVIDIA?'' A system succeeds only if it returns the requested values correctly and completely.

\textbf{Cost.} We compute cost as the sum of input and output tokens multiplied by model-specific rates. Text-to-SQL incurs substantially higher and more variable costs than FinAI Data Assistant across models, while the latter remains negligible in comparison.

\textbf{Latency.} We measure total time to completion. For FinAI Data Assistant, this includes function-call inference and linking-function execution (including SQL). For text-to-SQL, it includes SQL generation and execution. On a log scale, text-to-SQL is consistently slower.

\textbf{Accuracy.} Within our test suite, FinAI Data Assistant completed all cases successfully, whereas text-to-SQL exhibited failures across models. While our tasks are intentionally practical rather than adversarial, the results indicate that text-to-SQL is more error-prone than our function-calling approach under typical usage.

In sum, FinAI Data Assistant delivers strong practical performance relative to text-to-SQL in our evaluations.

\section{Conclusion}
We introduced \textbf{FinAI Data Assistant}, an LLM-driven system for natural-language access to financial databases that leverages the OpenAI Function Calling API and a small library of parameterized SQL templates. Our experiments support three conclusions. First, LLMs alone cannot reliably reconstruct time-varying financial values from short histories and show look-ahead bias primarily for prices, underscoring the need for explicit database integration (RQ1). Second, OpenAI LLMs exhibit near-perfect ticker mapping for NASDAQ-100 and high accuracy for S\&P~500 companies, enabling dependable entity resolution in common workflows (RQ2). Third, the function-calling approach achieves lower cost and latency with higher reliability than a text-to-SQL baseline on our task suite (RQ3).

\begin{acks}
This research was supported by the \textbf{Brian Impact Foundation}, a non-profit organization dedicated to the advancement of science and technology for all.

This document was prepared by the \textbf{AI Solution Team} at \textbf{Mirae Asset Global Investments} for informational purposes only. \textbf{Mirae Asset Global Investments} makes no representation or warranty as to the accuracy, completeness, or reliability of the information contained herein and accepts no liability for any consequences arising from its use. This document does not constitute investment advice, a recommendation, or an offer to buy or sell any financial products, and should not be relied upon as a basis for investment decisions.
\end{acks}

\bibliographystyle{ACM-Reference-Format}
\bibliography{reference}


\begin{thebibliography}{31}


\ifx \showCODEN    \undefined \def \showCODEN     #1{\unskip}     \fi
\ifx \showISBNx    \undefined \def \showISBNx     #1{\unskip}     \fi
\ifx \showISBNxiii \undefined \def \showISBNxiii  #1{\unskip}     \fi
\ifx \showISSN     \undefined \def \showISSN      #1{\unskip}     \fi
\ifx \showLCCN     \undefined \def \showLCCN      #1{\unskip}     \fi
\ifx \shownote     \undefined \def \shownote      #1{#1}          \fi
\ifx \showarticletitle \undefined \def \showarticletitle #1{#1}   \fi
\ifx \showURL      \undefined \def \showURL       {\relax}        \fi
\providecommand\bibfield[2]{#2}
\providecommand\bibinfo[2]{#2}
\providecommand\natexlab[1]{#1}
\providecommand\showeprint[2][]{arXiv:#2}

\bibitem[Crane et~al\mbox{.}(2025)]%
        {crane2025totalrecall}
\bibfield{author}{\bibinfo{person}{Leland~D. Crane}, \bibinfo{person}{Akhil Karra}, {and} \bibinfo{person}{Paul~E. Soto}.} \bibinfo{year}{2025}\natexlab{}.
\newblock \bibinfo{booktitle}{\emph{Total Recall? Evaluating the Macroeconomic Knowledge of Large Language Models}}.
\newblock \bibinfo{type}{Finance and Economics Discussion Series (FEDS)} 2025-044. \bibinfo{institution}{Board of Governors of the Federal Reserve System}.
\newblock
\href{https://doi.org/10.17016/FEDS.2025.044}{doi:\nolinkurl{10.17016/FEDS.2025.044}}


\bibitem[Glasserman and Lin(2024)]%
        {glasserman2024lookahead}
\bibfield{author}{\bibinfo{person}{Paul Glasserman} {and} \bibinfo{person}{Caden Lin}.} \bibinfo{year}{2024}\natexlab{}.
\newblock \showarticletitle{Assessing Look-Ahead Bias in Stock Return Predictions Generated by GPT Sentiment Analysis}.
\newblock \bibinfo{journal}{\emph{The Journal of Financial Data Science}} \bibinfo{volume}{6}, \bibinfo{number}{1} (\bibinfo{year}{2024}), \bibinfo{pages}{25--42}.
\newblock
\href{https://doi.org/10.3905/jfds.2023.1.143}{doi:\nolinkurl{10.3905/jfds.2023.1.143}}


\bibitem[Gruver et~al\mbox{.}(2023)]%
        {gruver2023zeroshot}
\bibfield{author}{\bibinfo{person}{Nate Gruver}, \bibinfo{person}{Marc Finzi}, \bibinfo{person}{Shikai Qiu}, {and} \bibinfo{person}{Andrew~Gordon Wilson}.} \bibinfo{year}{2023}\natexlab{}.
\newblock \showarticletitle{Large Language Models Are Zero-Shot Time Series Forecasters}. In \bibinfo{booktitle}{\emph{Advances in Neural Information Processing Systems (NeurIPS)}}.
\newblock
\urldef\tempurl%
\url{https://arxiv.org/abs/2310.07820}
\showURL{%
\tempurl}
\newblock
\shownote{arXiv:2310.07820}.


\bibitem[Hong et~al\mbox{.}(2024)]%
        {hong2024nextgen}
\bibfield{author}{\bibinfo{person}{Zijin Hong}, \bibinfo{person}{Zheng Yuan}, \bibinfo{person}{Qinggang Zhang}, \bibinfo{person}{Hao Chen}, \bibinfo{person}{Junnan Dong}, \bibinfo{person}{Feiran Huang}, {and} \bibinfo{person}{Xiao Huang}.} \bibinfo{year}{2024}\natexlab{}.
\newblock \showarticletitle{Next-Generation Database Interfaces: A Survey of LLM-based Text-to-SQL}.
\newblock \bibinfo{journal}{\emph{arXiv preprint}} (\bibinfo{year}{2024}).
\newblock
\showeprint[arxiv]{2406.08426}~[cs.DB]
\urldef\tempurl%
\url{https://arxiv.org/abs/2406.08426}
\showURL{%
\tempurl}


\bibitem[Kim et~al\mbox{.}(2025)]%
        {kim2025guruagents}
\bibfield{author}{\bibinfo{person}{Yejin Kim}, \bibinfo{person}{Youngbin Lee}, \bibinfo{person}{Juhyeong Kim}, {and} \bibinfo{person}{Yongjae Lee}.} \bibinfo{year}{2025}\natexlab{}.
\newblock \bibinfo{title}{GuruAgents: Emulating Wise Investors with Prompt-Guided LLM Agents}.
\newblock
\showeprint[arxiv]{2510.01664}~[cs.AI]


\bibitem[Lee et~al\mbox{.}(2025)]%
        {lee2025llmenhanced}
\bibfield{author}{\bibinfo{person}{Youngbin Lee}, \bibinfo{person}{Yejin Kim}, \bibinfo{person}{Juhyeong Kim}, \bibinfo{person}{Suin Kim}, {and} \bibinfo{person}{Yongjae Lee}.} \bibinfo{year}{2025}\natexlab{}.
\newblock \bibinfo{title}{LLM-Enhanced Black-Litterman Portfolio Optimization}.
\newblock
\showeprint[arxiv]{2504.14345}~[q-fin.PM]


\bibitem[Li et~al\mbox{.}(2023)]%
        {Li2023RESDSQL}
\bibfield{author}{\bibinfo{person}{Haoyang Li}, \bibinfo{person}{Jing Zhang}, \bibinfo{person}{Cuiping Li}, {and} \bibinfo{person}{Hong Chen}.} \bibinfo{year}{2023}\natexlab{}.
\newblock \showarticletitle{RESDSQL: Decoupling Schema Linking and Skeleton Parsing for Text-to-SQL}. In \bibinfo{booktitle}{\emph{Proceedings of the AAAI Conference on Artificial Intelligence}}, Vol.~\bibinfo{volume}{37}. \bibinfo{pages}{13067--13075}.
\newblock
\urldef\tempurl%
\url{https://ojs.aaai.org/index.php/AAAI/article/view/26535}
\showURL{%
\tempurl}


\bibitem[Lin et~al\mbox{.}(2020)]%
        {Lin2020BRIDGE}
\bibfield{author}{\bibinfo{person}{Xi~Victoria Lin}, \bibinfo{person}{Richard Socher}, {and} \bibinfo{person}{Caiming Xiong}.} \bibinfo{year}{2020}\natexlab{}.
\newblock \showarticletitle{Bridging Textual and Tabular Data for Cross-Domain Text-to-SQL Semantic Parsing}. In \bibinfo{booktitle}{\emph{Findings of the Association for Computational Linguistics: EMNLP 2020}}. \bibinfo{publisher}{Association for Computational Linguistics}.
\newblock
\href{https://doi.org/10.18653/v1/2020.findings-emnlp.438}{doi:\nolinkurl{10.18653/v1/2020.findings-emnlp.438}}


\bibitem[Liu et~al\mbox{.}(2025)]%
        {liu2025exante}
\bibfield{author}{\bibinfo{person}{Yachuan Liu}, \bibinfo{person}{Xiaochun Wei}, \bibinfo{person}{Lin Shi}, \bibinfo{person}{Xinnuo Li}, \bibinfo{person}{Bohan Zhang}, \bibinfo{person}{Paramveer Dhillon}, {and} \bibinfo{person}{Qiaozhu Mei}.} \bibinfo{year}{2025}\natexlab{}.
\newblock \showarticletitle{ExAnte: A Benchmark for Ex-Ante Inference in Large Language Models}.
\newblock \bibinfo{journal}{\emph{arXiv preprint}} (\bibinfo{year}{2025}).
\newblock
\showeprint[arxiv]{2505.19533}~[cs.LG]
\href{https://doi.org/10.48550/arXiv.2505.19533}{doi:\nolinkurl{10.48550/arXiv.2505.19533}}


\bibitem[Lopez-Lira et~al\mbox{.}(2025)]%
        {lopezlira2025memorization}
\bibfield{author}{\bibinfo{person}{Juan~Felipe Lopez-Lira}, \bibinfo{person}{Yuexi Tang}, {and} \bibinfo{person}{Lin Zhu}.} \bibinfo{year}{2025}\natexlab{}.
\newblock \showarticletitle{The Memorization Problem: Can We Trust LLMs’ Economic Forecasts?}
\newblock \bibinfo{journal}{\emph{arXiv preprint}} (\bibinfo{year}{2025}).
\newblock
\showeprint[arxiv]{2504.14765}~[econ.GN]
\urldef\tempurl%
\url{https://arxiv.org/abs/2504.14765}
\showURL{%
\tempurl}


\bibitem[OpenAI(2023)]%
        {OpenAI2023FunctionCalling}
\bibfield{author}{\bibinfo{person}{OpenAI}.} \bibinfo{year}{2023}\natexlab{}.
\newblock \bibinfo{title}{Function calling with the OpenAI API}.
\newblock \bibinfo{howpublished}{\url{https://platform.openai.com/docs/guides/function-calling}}.
\newblock
\newblock
\shownote{Accessed 2025-08-31}.


\bibitem[Pham and Cunningham(2024)]%
        {pham2024basechatgpt}
\bibfield{author}{\bibinfo{person}{Van Pham} {and} \bibinfo{person}{Scott Cunningham}.} \bibinfo{year}{2024}\natexlab{}.
\newblock \showarticletitle{Can Base ChatGPT be Used for Forecasting without Additional Optimization?}
\newblock \bibinfo{journal}{\emph{arXiv preprint}} (\bibinfo{year}{2024}).
\newblock
\showeprint[arxiv]{2404.07396}~[econ.GN]
\href{https://doi.org/10.48550/arXiv.2404.07396}{doi:\nolinkurl{10.48550/arXiv.2404.07396}}


\bibitem[Pourreza and Rafiei(2023)]%
        {Pourreza2023DINSQL}
\bibfield{author}{\bibinfo{person}{Mohammadreza Pourreza} {and} \bibinfo{person}{Davood Rafiei}.} \bibinfo{year}{2023}\natexlab{}.
\newblock \showarticletitle{DIN-SQL: Decomposed In-Context Learning of Text-to-SQL with Self-Correction}. In \bibinfo{booktitle}{\emph{Advances in Neural Information Processing Systems 36 (NeurIPS 2023)}}.
\newblock
\urldef\tempurl%
\url{https://proceedings.neurips.cc/paper_files/paper/2023/hash/72223cc66f63ca1aa59edaec1b3670e6-Abstract-Conference.html}
\showURL{%
\tempurl}


\bibitem[Prasad et~al\mbox{.}(2024)]%
        {Prasad2024Gorilla}
\bibfield{author}{\bibinfo{person}{Ishita Prasad}, \bibinfo{person}{Aman Madaan}, \bibinfo{person}{Shishir~G. Patil}, \bibinfo{person}{Mayee~F. Chen}, \bibinfo{person}{Percy Liang}, \bibinfo{person}{Joseph~E. Gonz\'alez}, {and} \bibinfo{person}{Ion Stoica}.} \bibinfo{year}{2024}\natexlab{}.
\newblock \showarticletitle{Gorilla: Large Language Model Connected with Massive APIs}. In \bibinfo{booktitle}{\emph{Advances in Neural Information Processing Systems 37 (NeurIPS 2024)}}.
\newblock
\urldef\tempurl%
\url{https://proceedings.neurips.cc/paper_files/paper/2024/file/e4c61f578ff07830f5c37378dd3ecb0d-Paper-Conference.pdf}
\showURL{%
\tempurl}


\bibitem[Quamar et~al\mbox{.}(2022)]%
        {quamar2022nli2d}
\bibfield{author}{\bibinfo{person}{Abdul Quamar}, \bibinfo{person}{Vasilis Efthymiou}, \bibinfo{person}{Chuan Lei}, {and} \bibinfo{person}{Fatma {\"O}zcan}.} \bibinfo{year}{2022}\natexlab{}.
\newblock \showarticletitle{Natural Language Interfaces to Data}.
\newblock \bibinfo{journal}{\emph{Foundations and Trends in Databases}} \bibinfo{volume}{11}, \bibinfo{number}{4} (\bibinfo{year}{2022}), \bibinfo{pages}{319--414}.
\newblock
\showISBNx{978-1-63828-028-6}
\href{https://doi.org/10.1561/1900000078}{doi:\nolinkurl{10.1561/1900000078}}


\bibitem[Rubin and Berant(2021)]%
        {Rubin2021SmBoP}
\bibfield{author}{\bibinfo{person}{Ohad Rubin} {and} \bibinfo{person}{Jonathan Berant}.} \bibinfo{year}{2021}\natexlab{}.
\newblock \showarticletitle{SmBoP: Semi-Autoregressive Bottom-Up Semantic Parsing}. In \bibinfo{booktitle}{\emph{Proceedings of the 2021 Conference of the North American Chapter of the Association for Computational Linguistics}}. \bibinfo{publisher}{Association for Computational Linguistics}.
\newblock
\urldef\tempurl%
\url{https://aclanthology.org/2021.naacl-main.29/}
\showURL{%
\tempurl}


\bibitem[Satterthwaite(1946)]%
        {satterthwaite1946approximate}
\bibfield{author}{\bibinfo{person}{Frank~E. Satterthwaite}.} \bibinfo{year}{1946}\natexlab{}.
\newblock \showarticletitle{An Approximate Distribution of Estimates of Variance Components}.
\newblock \bibinfo{journal}{\emph{Biometrics Bulletin}} \bibinfo{volume}{2}, \bibinfo{number}{6} (\bibinfo{year}{1946}), \bibinfo{pages}{110--114}.
\newblock
\href{https://doi.org/10.2307/3002019}{doi:\nolinkurl{10.2307/3002019}}


\bibitem[Schick et~al\mbox{.}(2023)]%
        {Schick2023Toolformer}
\bibfield{author}{\bibinfo{person}{Timo Schick}, \bibinfo{person}{Jane Dwivedi-Yu}, \bibinfo{person}{Roberto Dess\`{\i}}, \bibinfo{person}{Roberta Raileanu}, \bibinfo{person}{Maria Lomeli}, \bibinfo{person}{Eric Hambro}, \bibinfo{person}{Luke Zettlemoyer}, \bibinfo{person}{Nicola Cancedda}, {and} \bibinfo{person}{Thomas Scialom}.} \bibinfo{year}{2023}\natexlab{}.
\newblock \showarticletitle{Toolformer: Language Models Can Teach Themselves to Use Tools}. In \bibinfo{booktitle}{\emph{Advances in Neural Information Processing Systems 36 (NeurIPS 2023)}}.
\newblock
\urldef\tempurl%
\url{https://proceedings.neurips.cc/paper_files/paper/2023/file/d842425e4bf79ba039352da0f658a906-Paper-Conference.pdf}
\showURL{%
\tempurl}


\bibitem[Scholak et~al\mbox{.}(2021)]%
        {Scholak2021PICARD}
\bibfield{author}{\bibinfo{person}{Frederik Scholak}, \bibinfo{person}{Nathan Schucher}, {and} \bibinfo{person}{Dzmitry Bahdanau}.} \bibinfo{year}{2021}\natexlab{}.
\newblock \showarticletitle{PICARD: Parsing Incrementally for Constrained Auto-Regressive Decoding for Semantic Parsing}.
\newblock \bibinfo{journal}{\emph{arXiv preprint arXiv:2109.05093}} (\bibinfo{year}{2021}).
\newblock
\urldef\tempurl%
\url{https://arxiv.org/abs/2109.05093}
\showURL{%
\tempurl}


\bibitem[Shi et~al\mbox{.}(2025)]%
        {shi2025llmtexttosqlsurvey}
\bibfield{author}{\bibinfo{person}{Liang Shi}, \bibinfo{person}{Zhengju Tang}, \bibinfo{person}{Nan Zhang}, \bibinfo{person}{Xiaotong Zhang}, {and} \bibinfo{person}{Zhi Yang}.} \bibinfo{year}{2025}\natexlab{}.
\newblock \showarticletitle{A Survey on Employing Large Language Models for Text-to-SQL Tasks}.
\newblock \bibinfo{journal}{\emph{arXiv preprint arXiv:2407.15186}} (\bibinfo{year}{2025}).
\newblock
\urldef\tempurl%
\url{https://arxiv.org/abs/2407.15186}
\showURL{%
\tempurl}


\bibitem[Wang et~al\mbox{.}(2020)]%
        {Wang2020RATSQL}
\bibfield{author}{\bibinfo{person}{Bailin Wang}, \bibinfo{person}{Richard Shin}, \bibinfo{person}{Xiaodong Liu}, \bibinfo{person}{Oleksandr Polozov}, {and} \bibinfo{person}{Matthew Richardson}.} \bibinfo{year}{2020}\natexlab{}.
\newblock \showarticletitle{RAT-SQL: Relation-Aware Schema Encoding and Linking for Text-to-SQL Parsers}. In \bibinfo{booktitle}{\emph{Proceedings of the 58th Annual Meeting of the Association for Computational Linguistics}}. \bibinfo{publisher}{Association for Computational Linguistics}.
\newblock
\urldef\tempurl%
\url{https://aclanthology.org/2020.acl-main.677/}
\showURL{%
\tempurl}


\bibitem[Welch(1947)]%
        {welch1947generalization}
\bibfield{author}{\bibinfo{person}{B.~L. Welch}.} \bibinfo{year}{1947}\natexlab{}.
\newblock \showarticletitle{The Generalization of "Student's" Problem when Several Different Population Variances are Involved}.
\newblock \bibinfo{journal}{\emph{Biometrika}} \bibinfo{volume}{34}, \bibinfo{number}{1--2} (\bibinfo{year}{1947}), \bibinfo{pages}{28--35}.
\newblock
\href{https://doi.org/10.1093/biomet/34.1-2.28}{doi:\nolinkurl{10.1093/biomet/34.1-2.28}}


\bibitem[Wu et~al\mbox{.}(2023)]%
        {Wu2023BloombergGPT}
\bibfield{author}{\bibinfo{person}{Shijie Wu}, \bibinfo{person}{Ozan Irsoy}, \bibinfo{person}{Steven Lu}, \bibinfo{person}{Vadim Dabravolski}, \bibinfo{person}{Mark Dredze}, \bibinfo{person}{Sebastian Gehrmann}, \bibinfo{person}{Prabhanjan Kambadur}, \bibinfo{person}{David Rosenberg}, {and} \bibinfo{person}{Gideon Mann}.} \bibinfo{year}{2023}\natexlab{}.
\newblock \showarticletitle{BloombergGPT: A Large Language Model for Finance}.
\newblock \bibinfo{journal}{\emph{arXiv preprint arXiv:2303.17564}} (\bibinfo{year}{2023}).
\newblock
\urldef\tempurl%
\url{https://arxiv.org/abs/2303.17564}
\showURL{%
\tempurl}


\bibitem[Xie et~al\mbox{.}(2022)]%
        {Xie2022UnifiedSKG}
\bibfield{author}{\bibinfo{person}{Tianbao Xie}, \bibinfo{person}{Chen~Henry Wu}, \bibinfo{person}{Peng Shi}, \bibinfo{person}{Ruiqi Zhong}, \bibinfo{person}{Torsten Scholak}, \bibinfo{person}{Michihiro Yasunaga}, \bibinfo{person}{Chien-Sheng Wu}, \bibinfo{person}{Ming Zhong}, \bibinfo{person}{Pengcheng Yin}, \bibinfo{person}{Sida~I. Wang}, \bibinfo{person}{Victor Zhong}, \bibinfo{person}{Bailin Wang}, \bibinfo{person}{Chengzu Li}, \bibinfo{person}{Connor Boyle}, \bibinfo{person}{Ansong Ni}, \bibinfo{person}{Ziyu Yao}, \bibinfo{person}{Dragomir Radev}, \bibinfo{person}{Caiming Xiong}, \bibinfo{person}{Lingpeng Kong}, \bibinfo{person}{Rui Zhang}, \bibinfo{person}{Noah~A. Smith}, \bibinfo{person}{Luke Zettlemoyer}, {and} \bibinfo{person}{Tao Yu}.} \bibinfo{year}{2022}\natexlab{}.
\newblock \showarticletitle{UnifiedSKG: Unifying and Multi-Tasking Structured Knowledge Grounding with Text-to-Text Language Models}. In \bibinfo{booktitle}{\emph{Proceedings of the 2022 Conference on Empirical Methods in Natural Language Processing}}.
\newblock
\urldef\tempurl%
\url{https://arxiv.org/abs/2201.05966}
\showURL{%
\tempurl}
\newblock
\shownote{arXiv:2201.05966}.


\bibitem[Xu et~al\mbox{.}(2017)]%
        {Xu2017SQLNet}
\bibfield{author}{\bibinfo{person}{Xiaojun Xu}, \bibinfo{person}{Chang Liu}, {and} \bibinfo{person}{Dawn Song}.} \bibinfo{year}{2017}\natexlab{}.
\newblock \showarticletitle{SQLNet: Generating Structured Queries from Natural Language Without Reinforcement Learning}.
\newblock \bibinfo{journal}{\emph{arXiv preprint arXiv:1711.04436}} (\bibinfo{year}{2017}).
\newblock
\urldef\tempurl%
\url{https://arxiv.org/abs/1711.04436}
\showURL{%
\tempurl}


\bibitem[Yang et~al\mbox{.}(2023)]%
        {Yang2023FinGPT}
\bibfield{author}{\bibinfo{person}{Hongyang~(Bruce) Yang}, \bibinfo{person}{Xiao{-}Yang Liu}, {and} \bibinfo{person}{Christina~Dan Wang}.} \bibinfo{year}{2023}\natexlab{}.
\newblock \showarticletitle{FinGPT: Open-Source Financial Large Language Models}.
\newblock \bibinfo{journal}{\emph{arXiv preprint arXiv:2306.06031}} (\bibinfo{year}{2023}).
\newblock
\urldef\tempurl%
\url{https://arxiv.org/abs/2306.06031}
\showURL{%
\tempurl}


\bibitem[Yao et~al\mbox{.}(2023)]%
        {Yao2023ReAct}
\bibfield{author}{\bibinfo{person}{Shunyu Yao}, \bibinfo{person}{Jeffrey Zhao}, \bibinfo{person}{Dian Yu}, \bibinfo{person}{Nan Du}, \bibinfo{person}{Peter~J. Liu}, {and} \bibinfo{person}{Karthik Narasimhan}.} \bibinfo{year}{2023}\natexlab{}.
\newblock \showarticletitle{ReAct: Synergizing Reasoning and Acting in Language Models}. In \bibinfo{booktitle}{\emph{International Conference on Learning Representations (ICLR)}}.
\newblock
\urldef\tempurl%
\url{https://arxiv.org/abs/2210.03629}
\showURL{%
\tempurl}


\bibitem[Yu et~al\mbox{.}(2019)]%
        {Yu2019CoSQL}
\bibfield{author}{\bibinfo{person}{Tao Yu}, \bibinfo{person}{Rui Zhang}, \bibinfo{person}{He~Yang Er}, \bibinfo{person}{Suyi Li}, \bibinfo{person}{Eric Xue}, \bibinfo{person}{Bo Pang}, \bibinfo{person}{Xi~Victoria Lin}, \bibinfo{person}{Yi~Chern Tan}, \bibinfo{person}{Tianze Shi}, \bibinfo{person}{Zihan Li}, \bibinfo{person}{Youxuan Jiang}, \bibinfo{person}{Michihiro Yasunaga}, \bibinfo{person}{Sungrok Shim}, \bibinfo{person}{Tao Chen}, \bibinfo{person}{Alexander Fabbri}, \bibinfo{person}{Zifan Li}, \bibinfo{person}{Luyao Chen}, \bibinfo{person}{Yuwen Zhang}, \bibinfo{person}{Shreya Dixit}, \bibinfo{person}{Vincent Zhang}, \bibinfo{person}{Caiming Xiong}, \bibinfo{person}{Richard Socher}, \bibinfo{person}{Walter~S. Lasecki}, {and} \bibinfo{person}{Dragomir Radev}.} \bibinfo{year}{2019}\natexlab{}.
\newblock \showarticletitle{CoSQL: A Conversational Text-to-SQL Challenge Towards Cross-Domain Natural Language Interfaces to Databases}. In \bibinfo{booktitle}{\emph{Proceedings of the 2019 Conference on Empirical Methods in Natural Language Processing}}. \bibinfo{publisher}{Association for Computational Linguistics}.
\newblock
\urldef\tempurl%
\url{https://aclanthology.org/D19-1204/}
\showURL{%
\tempurl}


\bibitem[Yu et~al\mbox{.}(2018)]%
        {Yu2018Spider}
\bibfield{author}{\bibinfo{person}{Tao Yu}, \bibinfo{person}{Rui Zhang}, \bibinfo{person}{Kai Yang}, \bibinfo{person}{Michihiro Yasunaga}, \bibinfo{person}{Dongxu Wang}, \bibinfo{person}{Zifan Li}, \bibinfo{person}{James Ma}, \bibinfo{person}{Irene Li}, \bibinfo{person}{Qingning Yao}, \bibinfo{person}{Shanelle Roman}, \bibinfo{person}{Zilin Zhang}, {and} \bibinfo{person}{Dragomir Radev}.} \bibinfo{year}{2018}\natexlab{}.
\newblock \showarticletitle{Spider: A Large-Scale Human-Labeled Dataset for Complex and Cross-Domain Semantic Parsing and Text-to-SQL}. In \bibinfo{booktitle}{\emph{Proceedings of the 2018 Conference on Empirical Methods in Natural Language Processing}}. \bibinfo{publisher}{Association for Computational Linguistics}, \bibinfo{address}{Brussels, Belgium}.
\newblock
\urldef\tempurl%
\url{https://aclanthology.org/D18-1425/}
\showURL{%
\tempurl}


\bibitem[Zhong et~al\mbox{.}(2017)]%
        {Zhong2017Seq2SQL}
\bibfield{author}{\bibinfo{person}{Victor Zhong}, \bibinfo{person}{Caiming Xiong}, {and} \bibinfo{person}{Richard Socher}.} \bibinfo{year}{2017}\natexlab{}.
\newblock \showarticletitle{Seq2SQL: Generating Structured Queries from Natural Language using Reinforcement Learning}.
\newblock \bibinfo{journal}{\emph{arXiv preprint arXiv:1709.00103}} (\bibinfo{year}{2017}).
\newblock
\urldef\tempurl%
\url{https://arxiv.org/abs/1709.00103}
\showURL{%
\tempurl}


\bibitem[Zhu et~al\mbox{.}(2024)]%
        {zhu2024llmtexttosqlsurvey}
\bibfield{author}{\bibinfo{person}{Xiaohu Zhu}, \bibinfo{person}{Qian Li}, \bibinfo{person}{Lizhen Cui}, {and} \bibinfo{person}{Yongkang Liu}.} \bibinfo{year}{2024}\natexlab{}.
\newblock \showarticletitle{Large Language Model Enhanced Text-to-SQL Generation: A Survey}.
\newblock \bibinfo{journal}{\emph{arXiv preprint}} (\bibinfo{year}{2024}).
\newblock
\showeprint[arxiv]{2410.06011}~[cs.DB]
\urldef\tempurl%
\url{https://arxiv.org/abs/2410.06011}
\showURL{%
\tempurl}


\end{thebibliography}

\clearpage
\appendix

\section{Test cases for Database Query Performance (RQ3) experiment.}

\begin{tcolorbox}[title=test cases,
  colback=green!10,   
  colframe=green!40!black, 
  boxrule=0.6pt,       
  arc=2mm,             
  breakable            
]
\small
Return NVDA's stock price for the past 1 days.\\
Return Walmart's stock price for the past 1 days.\\
Return MSFT’s stock price for the past 5 days.\\
Return Meta’s stock price for the past 5 days.\\
Return Coca-Cola's revenue for 1 quarter.\\
Return TSLA's total asset for 5 quarter.\\
Return Costco's yearly net income for 1 year.\\
Return AMZN's yearly interest expense for 5 year.\\
Return AAPL's stock price for the past 3 days.\\
Return GOOGL’s stock price for the past 10 days.\\
Return TSLA's stock price for the past 8 weeks.\\
Return AMZN's stock price for the past 12 months.\\
Show JPM’s stock price between 2024-11-01 and 2024-12-31.\\
Return NVDA’s stock price for the past 1 year with weekly periodicity.\\
Return META’s stock price for the past 5 days\\
Return MSFT’s stock price for the past 6 months.\\
Return Apple’s revenue for the last 4 quarters.\\
Return Alphabet’s net income for 2 quarters.\\
Return Amazon’s total assets for 3 quarters.\\
Return Tesla’s total liabilities for 4 quarters.\\
Return Microsoft’s operating income for 5 quarters.\\
Return Nvidia’s R\&D expense for 4 quarters.\\
Return Walmart’s SG\&A expense for 3 quarters.\\
Return Coca-Cola’s gross profit for 2 quarters.\\
Return Costco’s operating cash flow for 4 quarters.\\
Return Meta’s capital expenditures for 4 quarters.\\
Return the income statement for AAPL for the last 2 quarters.\\
Return the balance sheet for MSFT for the last 1 quarter.\\
Return the cash flow statement for AMZN for the last 3 quarters.\\
Return NVDA’s income statement between 2023 Q2 and 2024 Q1.\\
Return TSLA’s balance sheet between 2022 Q4 and 2023 Q3.\\
Return WMT’s cash flow statement for 2021 Q1 to 2022 Q2.\\
Return META’s yearly net income for the past 3 years.\\
Return ORCL’s yearly revenue for the past 5 years.\\
Return NFLX’s yearly interest expense for the past 2 years.\\
Return PEP’s yearly dividends paid for the past 3 years.\\
Return INTC’s yearly total asset for the past 5 years.\\
Return IBM’s yearly EPS (diluted) for the past 3 years.\\
Return AAPL's stock price for the past 7 days.\\
Return MSFT’s stock price for the past 14 days.\\
Return TSLA’s stock price for the past 30 days.\\
Return META’s stock price for the past 90 days.\\
Return AMZN’s stock price for the past 180 days.\\
Return NVDA’s stock price for the past 2 years.\\
Return GOOGL’s stock price for the past 3 years.\\
Return NFLX’s stock price for the past 5 days.\\
Return ORCL’s stock price for the past 10 days.\\
Return IBM’s stock price for the past 2 weeks.\\
Return INTC’s stock price for the past 6 months.\\
Return META’s stock price for the past 1 year.\\
Return JPM’s stock price between 2024-01-01 and 2024-03-31.\\
Return PEP’s stock price between 2023-05-01 and 2023-06-15.\\
Return WMT’s stock price for the past 2 years with monthly periodicity.\\
Return KO’s stock price for the past 4 years with yearly periodicity.\\
Return MSFT’s daily closing price for the past 20 days.\\
Return TSLA’s stock price for the past 15 days.\\
Return AAPL’s stock price from 2023-01-01 to 2023-06-30.\\
Return AMZN’s stock price for the past 3 months.\\
Return AAPL’s quarterly revenue for the past 6 quarters.\\
Return MSFT’s quarterly net income for the past 8 quarters.\\
Return TSLA’s quarterly operating income for the past 10 quarters.\\
Return NVDA’s quarterly gross profit for the past 12 quarters.\\
Return META’s quarterly cost of revenue for the past 4 quarters.\\
Return AMZN’s quarterly EPS (basic) for the past 6 quarters.\\
Return WMT’s quarterly EPS (diluted) for the past 8 quarters.\\
Return KO’s quarterly total liabilities for the past 4 quarters.\\
Return PEP’s quarterly total equity for the past 6 quarters.\\
Return INTC’s quarterly research and development expense for 5 quarters.\\
Return NFLX’s quarterly SG\&A expense for 3 quarters.\\
Return ORCL’s quarterly interest expense for 4 quarters.\\
Return IBM’s quarterly income tax expense for 2 quarters.\\
Return META’s quarterly total assets for 5 quarters.\\
Return JPM’s quarterly net interest income for 3 quarters.\\
Return BAC’s quarterly provisions for credit losses for 4 quarters.\\
Return GS’s quarterly revenue for 6 quarters.\\
Return MS’s quarterly net income for 8 quarters.\\
Return C’s quarterly EPS for 5 quarters.\\
Return V’s quarterly operating margin for 3 quarters.\\
Return AAPL’s yearly revenue for the past 10 years.\\
Return MSFT’s yearly operating income for the past 7 years.\\
Return TSLA’s yearly net income for the past 5 years.\\
Return NVDA’s yearly R\&D expense for the past 8 years.\\
Return META’s yearly capital expenditures for the past 6 years.\\
Return AMZN’s yearly total asset for the past 4 years.\\
Return WMT’s yearly SG\&A expense for the past 3 years.\\
Return KO’s yearly net income for the past 12 years.\\
Return PEP’s yearly gross profit for the past 9 years.\\
Return INTC’s yearly total assets for the past 15 years.\\
Return NFLX’s yearly operating cash flow for the past 7 years.\\
Return ORCL’s yearly dividends paid for the past 5 years.\\
Return IBM’s yearly EPS (basic) for the past 8 years.\\
Return META’s yearly gross profit for the past 6 years.\\
Return JPM’s yearly interest income for the past 4 years.\\
Return BAC’s yearly interest expense for the past 5 years.\\
Return GS’s yearly operating revenue for the past 7 years.\\
Return MS’s yearly income before tax for the past 6 years.\\
Return C’s yearly net income for the past 3 years.\\
Return V’s yearly transaction revenue for the past 5 years.\\
Return the income statement for TSLA for the last 3 quarters.\\
Return the balance sheet for NVDA for the last 2 quarters.\\
Return the cash flow statement for META for the last 4 quarters.\\
Return AAPL’s income statement for the past 2 years.\\
Return MSFT’s balance sheet for the past 5 years.\\
Return AMZN’s cash flow statement for the past 3 years.\\
Return WMT’s income statement from 2020 to 2022.\\
Return KO’s balance sheet between 2018 and 2020.\\
Return PEP’s cash flow statement from 2021 to 2023.\\
Return NFLX’s income statement for 2022 Q1 to 2023 Q4.\\
Return ORCL’s balance sheet for 2019 to 2021.\\
Return IBM’s cash flow statement between 2020 and 2022.\\
\end{tcolorbox}

\end{document}